# Observation of quantum interference of optical transition pathways in Doppler-free two-photon spectroscopy and implications for precision measurements

Bubai Rahaman[1], Sid C. Wright[2], and Sourav Dutta[1,*]

[1]*Tata Institute of Fundamental Research, 1 Homi Bhabha Road, Colaba, Mumbai 400005, India*
[2]*Fritz-Haber-Institut der Max-Planck-Gesellschaft, Faradayweg 4-6, 14195 Berlin, Germany*

Doppler-free two-photon spectroscopy is a standard technique for precision measurement of transition frequencies of dipole-forbidden transitions. The accuracy of such measurements depends critically on fitting of the spectrum to an appropriate line shape model, often a Voigt profile which neglects the effect of quantum interference of optical transitions. Here, we report the observation of quantum interference of optical transition pathways in Doppler-free two-photon spectroscopy of the cesium *6S*–*7D* transitions. The quantum interference manifests itself as asymmetric line shapes of the hyperfine lines of the *7D* states, observed through spontaneous emission following excitation by a narrow-linewidth cw laser. The interference persists despite the lines being spectrally well-resolved. Ignoring the effect of quantum interference causes large systematic shifts in the determination of the line-centers, while accounting for it resolves the apparent line shift and enables the precise determination of hyperfine splitting in the *7D* states. We calculate the spectral line shape including the effect of quantum interference and show that it agrees with the experimental observations. Our results have implications for precision measurements of hyperfine splittings, isotope shifts and transition frequencies, including those of the hydrogen *S*–*S* and *S*–*D* transitions.

Precise measurements of transition frequencies are required to benchmark theories such as quantum electrodynamics [1,2], determine fundamental physical constants [3,4], test the temporal variations of the fine structure constant [5], search for dark matter [6], search for physics beyond the standard model through isotope-shift measurements [7–9] and improve the performance of atomic clocks [10]. One of the most straight-forward and elegant techniques for high precision measurement of transition frequencies relies on Doppler-free two-photon spectroscopy [11,12]. It has been used for several decades to measure the hydrogen *1S*–*2S* transition frequency resulting in a fractional uncertainty as low as $4.2 \times 10^{-15}$ [13]. More recently, the technique was used to measure the hydrogen *1S*–*3S* transition frequency [4,14] and combined with the *1S*–*2S* measurement to determine the Rydberg constant and the proton charge radius. The technique has also been used for the measurements of isotope shifts on the $6s^2\, ^1S_0 \rightarrow 5d6s\, ^1D_2$ transition in ytterbium [9] and hyperfine splitting (HFS) of the *S* and *D* states of several atoms [15]. This includes our recent measurements of the HFS in the $7d\,^2D_{3/2}$ and $7d\,^2D_{5/2}$ states of cesium [16,17] which is motivated by the possibility of measuring the Cs *S*–*D* parity-nonconserving amplitudes [18,19]. (Hereafter, we use the abbreviated notation $ns_{1/2}$, $np_J$ and $nd_J$ to denote the $ns\,^2S_{1/2}$, $np\,^2P_J$ and $nd\,^2D_J$ states in Cs, respectively. *F*, *F′* and *F″* denote hyperfine states).

These experiments include diligent characterization and reduction of systematic effects, for example, due to the ac Stark shift, collisional shift and the residual Doppler shift. However, what is often overlooked is the systematic line shifts caused by quantum interference (QI) when energy levels are closely spaced. Remarkably, such shifts manifest in the detection of spontaneous emission. In this article, we report the experimental observation of asymmetric lineshapes in Doppler-free two-photon spectroscopy of the cesium $6s_{1/2}$–$7d_J$ transitions arising from QI of optical transition pathways and show that it causes systematic line shifts even when the transition lines are spectrally well-resolved. We study both $7d_{3/2}$ and $7d_{5/2}$ states to establish the generality of the effect. Accounting for the QI allows us to determine the HFS of the $7d_J$ states with unprecedented accuracy and put bounds on the nuclear magnetic octupole moment (Ω).

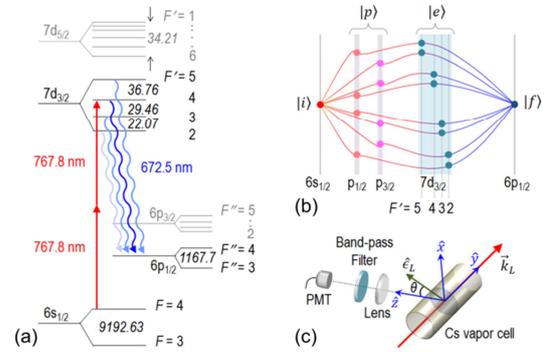

FIG. 1. Schematic representation of the experiment. **(a)** Partial energy level diagram (not to scale) of $^{133}$Cs. Numbers in italics are the HFS in MHz. Two-photon excitation of the $7d_{3/2}$ state and the detected decay channels are depicted. Probability amplitudes of exciting different *F′* levels are different (i.e. dependent on the laser detuning). Different shades of the wavy lines represent the different amplitudes of the possible decay channels. **(b)** Schematic representation of interfering quantum pathways, originating in state $|i\rangle$ and culminating in state $|f\rangle$ but accessing different intermediate states $|p\rangle$ and $|e\rangle$, resonantly or non-resonantly. **(c)** Sketch of the experimental setup. The laser with wave vector $\vec{k}_L$ propagates along the $\hat{y}$ direction, is linearly polarized along $\hat{\epsilon}_L$ in the $\hat{x} - \hat{z}$ plane and makes an angle $\theta$ with the $\hat{z}$ axis. The detection is along the $\hat{z}$ direction.

The QI arises when an atom in an initial state can be transferred to a final state via multiple undetected resonant and/or off-resonant intermediate states. The effect of QI was first observed in one-photon transitions [20,21], albeit in a different context. More recently, calculations [22] and observations [23–25] have shown that QI in one-photon transitions leads to systematic line shifts when the transition lines are overlapping or close in frequency. For two-photon transitions, the experimental observation was pending. Calculations, however, have highlighted the important implications of QI in the context of hydrogen 1$S$–3$S$ spectroscopy [26,27] and our analysis is largely based on Ref. [26] adapted for cw laser excitation.

The energy level diagram relevant to the experiment is shown in Fig. 1(a), the concept of QI is depicted in Fig. 1(b) and the sketch of the setup is shown in Fig. 1(c). The $6s_{1/2}(F = 3)$ and $6s_{1/2}(F = 4)$ states are equally populated in a Cs vapour cell but can be selectively excited to the $7d_{3/2}$ ($7d_{5/2}$) state using a narrow-linewidth cw lasers near 767.8 nm (767.2 nm). The different hyperfine levels $F'$ are excited by tuning the frequency of excitation laser using an acousto-optic modulator (AOM) in double-pass configuration [16,17]. The spontaneous emission at 672.5 nm (695 nm) due the $7d_{3/2} \to 6p_{1/2}$ ($7d_{5/2} \to 6p_{3/2}$) decay is recorded as laser frequency is tuned, generating the hyperfine spectrum for the $7d_{3/2}$ ($7d_{5/2}$) state. The chosen decay channel has the highest branching ratio ~0.65 (~0.77) [28] and the emitted light cannot be reabsorbed by the ground state Cs atoms resulting in high signal-to-noise ratio (SNR) of the order $10^3$. Additional details of the experiments are provided in Supplementary Material (SM) [29].

We show the experimentally measured spectra for the $7d_{3/2}$ ($7d_{5/2}$) state in Fig. 2 (Fig. 3) for different angles $\theta$ between the linearly polarized incident light and the detector direction. The panels (a) and (b) depict the spectra recorded when atoms are excited from the $6s_{1/2}(F = 3)$ and $6s_{1/2}(F = 4)$ levels, respectively. The leftmost, middle and rightmost panels represent spectra taken for $\theta \sim 0°\pm 2°$, $53°\pm 2°$ and $90°\pm 2°$, respectively. Each spectrum of the $7d_{3/2}$ ($7d_{5/2}$) state is fit to a combination of 4 (5) independent Voigt line shapes, from which we determine the 4 (5) line-centers of the individual peaks, 4 (5) Lorentzian widths, 4 (5) Gaussian widths, 4 (5) peak heights and 1 (1) background offset. The fit residual is shown below each spectrum. The HFS is obtained from the difference in the line-centers and is plotted in panel (c).

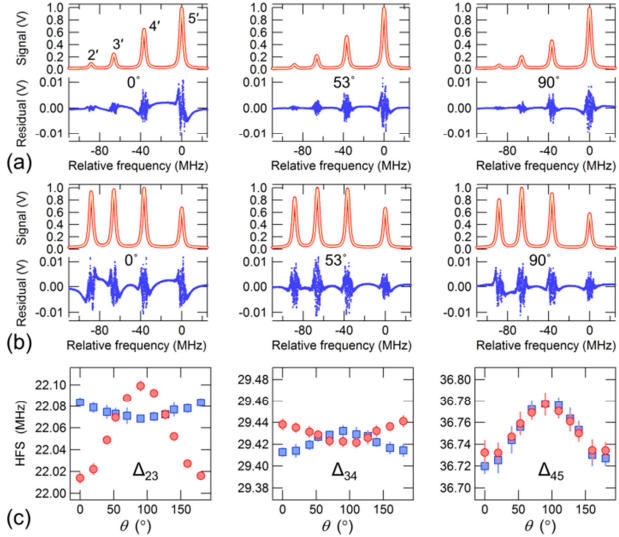

**FIG. 2. Spectroscopy of the $7d_{3/2}$ state. (a, b)** Spectra for the $6s_{1/2}(F = 3) \to 7d_{3/2}(F' = 2,3,4,5)$ and the $6s_{1/2}(F = 4) \to 7d_{3/2}(F' = 2,3,4,5)$ transitions, respectively, at three different values of $\theta$ viz. 0°, 53° and 90°. Red lines are the measured data and the superimposed white lines are the fits to a combination of four independent Voigt profiles, each of which is symmetric around the line-center. The residuals (blue dots) show the difference between the data points and the fitted function. The asymmetry in the residuals on either side of the line-center is most clearly visible in the $6s_{1/2}(F = 3) \to 7d_{3/2}(F' = 5)$ and the $6s_{1/2}(F = 4) \to 7d_{3/2}(F' = 2)$ lines for $\theta = 0°$. The asymmetry is reduced to noise level for $\theta = 53°$ and reappears for $\theta = 90°$. **(c)** The $7d_{3/2}$ state HFS $\Delta_{23}$, $\Delta_{34}$ and $\Delta_{45}$ obtained from the Voigt fits depend on $\theta$ due QI. Blue squares: excitation from the $6s_{1/2}(F = 3)$ state, red circles: excitation from the $6s_{1/2}(F = 4)$ state. The HFS for the two cases agree when $\theta \sim 54.7°$ (and equivalently also at $\theta \sim 125.3°$) where the QI vanishes.

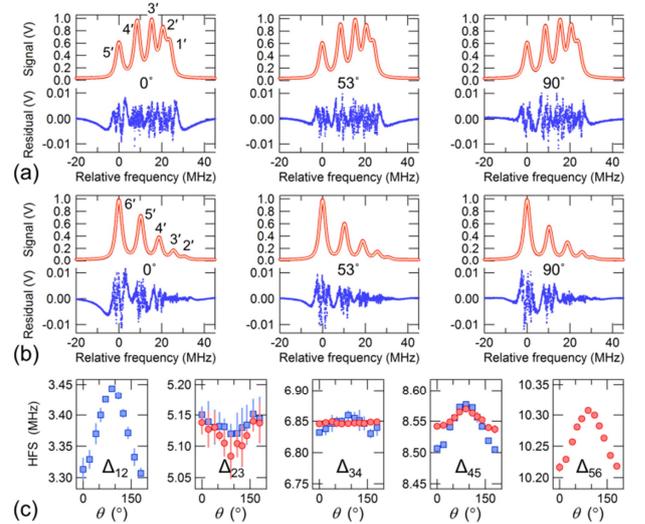

**FIG. 3. Spectroscopy of the $7d_{5/2}$ state.** The colour code and notations follow the conventions set in Fig. 2. **(a, b)** Spectra for the $6s_{1/2}(F = 3) \to 7d_{5/2}(F' = 1,2,3,4,5)$ and the $6s_{1/2}(F = 4) \to 7d_{5/2}(F' = 2,3,4,5,6)$ transitions. The asymmetry in the residuals are most clearly visible in the $6s_{1/2}(F = 3) \to 7d_{5/2}(F' = 5)$ and the $6s_{1/2}(F = 4) \to 7d_{5/2}(F' = 6)$ lines for $\theta = 0°$. **(c)** The $7d_{5/2}$ state HFS $\Delta_{12}$, $\Delta_{23}$, $\Delta_{34}$, $\Delta_{45}$ and $\Delta_{56}$ obtained from the Voigt fits depend on $\theta$.

The residuals have different signs on the two sides of the line-center for all angles other than $\theta \sim 53°$. The dispersive shaped pattern in the residuals provides striking evidence of asymmetry in the peaks. The dispersive pattern disappears at $\theta \sim 53°$ suggesting that the Voigt profile is the correct line shape near this angle. Furthermore, it is seen that the slope of the dispersive pattern changes sign at $\theta \sim 53°$. Another striking observation [Figs. 2(c), 3(c)] is that the measured $7d_J$ state HFS are different when experiments are performed by exciting atoms from the $6s_{1/2}(F=3)$ and $6s_{1/2}(F=4)$ levels, except when experiments are performed at $\theta \sim 53°$ i.e. near the "magic angle" $\theta_m = 54.7°$. Deviations as large as 100 kHz are seen. These observations, when compared with calculations (Figs. 4 and 5), prove the existence of QI of optical transition pathways as discussed below.

Stimulated absorption of the laser light followed by spontaneous emission transfers atoms in an initial $6s_{1/2}$ state $|i\rangle$ to the final $6p_J$ state $|f\rangle$ via several distinct quantum pathways which differ in the intermediate $7d_J$ states $|e\rangle$ accessed during the excitation and de-excitation processes [see Fig. 1(c)]. Following the perturbative treatment developed in Ref. [26], the second-order dipole matrix element for the $|i\rangle \to |e\rangle$ two-photon excitation process is

$$Q_{ei}^{\mu\nu} = \sum_p \frac{D_{ep}^\nu D_{pi}^\mu + D_{ep}^\mu D_{pi}^\nu}{\hbar(\omega_{pi} - \omega_L)},$$

where $D_{ep}^\nu$ ($D_{pi}^\mu$) is the dipole matrix elements between $|e\rangle$ and $|p\rangle$ ($|p\rangle$ and $|i\rangle$) with $|p\rangle$ denoting the intermediate $6p_J$ states accessible via 1-photon transitions, $\mu$ and $\nu$ are the incoming and retro-reflected laser polarizations (identical in our case) and $\omega_L$ is the laser frequency. The subsequent radiative decay $|e\rangle \to |f\rangle$ entails a dipole matrix element $D_{fe}^\eta$, where $\eta$ is the polarization of the emitted radiation. The emitted intensity takes the form

$$I^{\eta\mu\nu} \propto \sum_{if}\left|\sum_e \frac{D_{fe}^\eta Q_{ei}^{\mu\nu} H(\omega_{ef})}{\omega_{ei} - 2\omega_L - i\Gamma_e/2}\right|^2,$$

where $H(\omega_{ef})$ is proportional to the density of states and $\Gamma_e$ is the decay rate. Since the amplitudes of the various processes are added before taking the modulus square, cross terms are obtained in addition to the standard Lorentzian terms. The cross terms introduce an additional $\theta$−dependence to the fluorescence emission pattern. This $\theta$−dependence is over and above that arising from the dipole emission pattern alone, and causes asymmetry in the line shape. In general, the cross terms depend on $\theta$, specifically on $P_2(\cos\theta)$, but identically go to zero when $3\cos^2\theta - 1 = 0$ i.e. $\theta \equiv \theta_m = 54.7°$, in which case the line shape can be represented as a symmetric Lorentzian profile (or a symmetric Voigt profile, if transit time broadening and collisional broadening are included). In a real experiment, the solid angle $\Omega_c$ subtended by the imaging lens is non-zero. For small values of $\Omega_c$ (i.e. $\Omega_c \ll 4\pi$), as in our experiment, the interference term reduces in magnitude but is still appreciable. However, it becomes zero for $\Omega_c = 4\pi$. The expressions for the full quantum interference line shape for arbitrary $\theta$ and additional details on the model are provided in the SM [29].

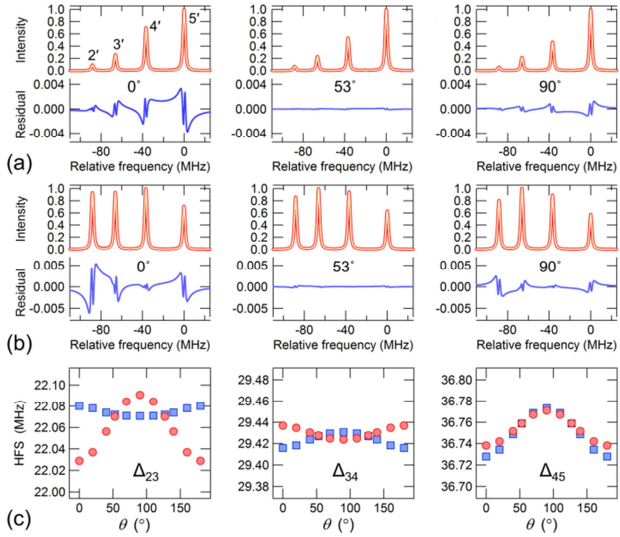

**FIG. 4. Simulated spectra of $7d_{3/2}$ state. (a, b)** The $6s_{1/2}(F=3) \to 7d_{3/2}(F'=2-5)$ and the $6s_{1/2}(F=4) \to 7d_{3/2}(F'=2-5)$ transitions, respectively. The red lines are the simulated data and the superimposed white lines are the fits to a combination of four independent Voigt profiles. The residuals (blue line) are asymmetric on either side of the line-center except when $\theta = 53°$, where the residuals tend to zero. The sign of the residuals change sign at $\theta \sim 54.7°$. **(c)** The $7d_{3/2}$ state HFS obtained from the Voigt fits vary with $\theta$ due to QI. The HFS when exciting from the $6s_{1/2}(F=3)$ state (blue squares) and the $6s_{1/2}(F=4)$ state (red circles) agree at $\theta \sim 54.7°$ (and equivalently also at $\theta \sim 125.3°$) where the QI vanishes.

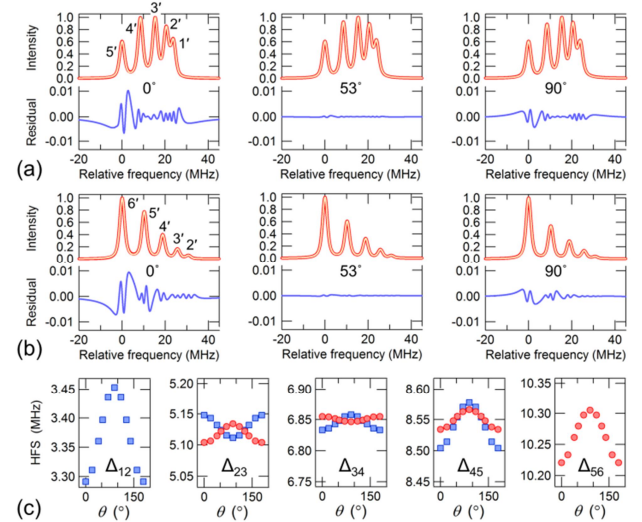

**FIG. 5. Simulated spectra of $7d_{5/2}$ state.** The color code and notations follow the conventions set in Fig. 4. **(a, b)** The $6s_{1/2}(F=3) \to 7d_{5/2}(F'=1-5)$ and the $6s_{1/2}(F=4) \to 7d_{5/2}(F'=2-6)$ transitions, respectively. **(c)** The $7d_{5/2}$ state HFS obtained from the Voigt fits.

**Table I: The hyperfine coupling constants determined in this work and a comparison with earlier reports.**

| | HCC | This work | Ref. [16] | Ref. [17] | Ref. [30] | Ref. [31] | Ref. [32] | Ref. [33] | Ref. [34] | Ref. [35] | Ref. [36] | Ref. [37] |
|---|---|---|---|---|---|---|---|---|---|---|---|---|
| | $A$ (MHz) | 7.3547(8) | 7.3509(9) | | 7.386(15) | 7.38(1) | 7.36(3) | 7.36(7) | 7.39(6) | 7.48 | 7.88 | 7.42(37) |
| $7d_{3/2}$ | $B$ (MHz) | −0.017(7) | −0.041(8) | | −0.18(16) | −0.18(10) | −0.1(2) | −0.88(87) | −0.19(18) | | | −0.0249(9) |
| | $C$ (kHz) | −0.3(4) | −0.03(53) | | | | | | | | | |
| | $A$ (MHz) | −1.7110(3) | | −1.7087(6) | −1.717(15) | | | −1.81(5) | −1.79(5) | −1.13 | −1.42 | −1.26(16) |
| $7d_{5/2}$ | $B$ (MHz) | −0.027(7) | | 0.050(14) | −0.18(52) | | | 1.01(1.06) | 1.05(29) | | | −0.0339(12) |
| | $C$ (kHz) | −0.02(80) | | 0.4(1.4) | | | | | | | | |

In Figs. 4(a,b) and 5(a,b), we plot the computed spectra taking into consideration the QI effect, detector geometry, solid angle and a Gaussian broadening to mimic the transit-time and collisional broadening seen in the experiment. The HFS are supplied as input (from the experimental measurements at $\theta \sim \theta_m$) and all other parameters are fixed at their respective theoretical values. The computed spectra reproduce the experimentally observed spectra (Figs. 2 and 3). We fit the computed spectra to Voigt line shapes in an identical manner as done for the experimental data. The fit residuals show the same features that are experimentally observed in Figs. 2 and 3. Moreover, the extracted HFS vs. $\theta$ plots [Figs. 4(c) and 5(c)] closely reproduce our experimental observations [Figs. 2(c) and 3(c)], not just qualitatively but also quantitatively. The experimental data [Figs. 2 and 3] and the computed spectra [Figs. 4 and 5] together provide the necessary evidence in support of the observation of QI of optical transition pathways in Doppler-free two photon spectroscopy and highlights that fitting to a Voigt line shape results in apparent line shifts.

Finally, we fit the experimental data to the QI line shape model and find that the dependence of HFS on $\theta$ is much reduced but not completely removed. The fitting parameters were: the line-centers, a common Lorentzian linewidth, a common Gaussian linewidth, an overall amplitude and an overall offset. All other parameters were fixed – the relative peak heights were fixed at their theoretical values; $\Omega_c$ and $\theta$ were fixed at their measured values. That the $\theta$-dependence of the HFS is reduced, despite a smaller number of free parameters in the fitting function, supports the efficacy of the QI model. It is difficult to accurately define the true fluorescence volume within the laser focus (see SM [29]), and we believe this explains the remnant $\theta$-dependence we observe.

We report the recommended values of HFS in Table 1S of the SM [29] along with the statistical uncertainties. We account for the systematic uncertainties arising from the ac Stark shift (~3 kHz), collisional shift (~5 kHz), Zeeman shift (<1 kHz) and second order Doppler shift (~0.5 kHz), as discussed in the SM [29]. The statistical and systematic uncertainties are added in quadrature to estimate the total uncertainty in the HFS and the hyperfine coupling constants (HCCs). We report the values of the HCCs in Table I. The expressions used to calculate the magnetic dipole ($A$), the electric quadrupole ($B$) and the magnetic octupole ($C$) coupling constant from the measured HFS are provided in the SM [29]. The recently reported corrections due to the second order effects [37,38] are included in our analysis.

We improve the precision of the HCCs by at least an order of magnitude compared to experimental reports from other groups [30–34]. Compared to our earlier works [16,17], the precision is marginally improved but, more importantly, the mean values have changed significantly on accounting for the quantum interference. Notably, using the calculated values of $C/\Omega = 0.0195$ (−0.0180) kHz/($\mu_N \times$b) for the $7d_{3/2}$ ($7d_{5/2}$) state [37] and our values of $C$, we determine $\Omega = -15 \pm 20$ ($1 \pm 44$) $\mu_N \times$b. While this puts bounds on the value of $\Omega$, the error bars are still too large to constrain nuclear model calculations. This indicates that higher precision experiments must be undertaken in the future. On the other hand, the experimentally determined values of $A$ and $B$ are precise but not in perfect agreement with the theoretical values [35–37] suggesting that more sophisticated theoretical calculations of HCCs for the $7d$ and other $nd$ states needs to be undertaken.

In summary, we show that the QI between optical transition pathways leads to asymmetric line shapes in Doppler-free two-photon spectroscopy. If unaccounted for, this causes apparent shifts of several tens of kHz in determination of line-centers of the Cs $7d_J$ states. Importantly, the interference effect is present although the lines are well-resolved and the laser has narrow linewidth. Therefore its influence must be considered in all other fluorescence-based two-photon spectroscopy experiments. The effect of QI vanishes at the magic angle of 54.7° between the laser polarization and the detector axis, thus providing a convenient practical alternative to the full QI model fitting.

**Acknowledgements**
We acknowledge funding from the Department of Atomic Energy, Government of India under Project Identification No. RTI4002. S.C.W. acknowledges useful discussions with S. Hofsäss and G. Meijer.

*sourav.dutta@tifr.res.in

# Supplementary information

## A. Experimental setup:

Figure 1S shows the schematic diagram of the experimental setup. The frequency of the laser is stabilized using an electronic feedback from the Doppler-free two-photon spectroscopy in Cs cell 1. The hyperfine spectra (e.g. Figs. 2 and 3 of the main article) are obtained by tuning the radio-frequency (rf) applied to AOM2 and recording the fluorescence from Cs cell 2 on a PMT. The AOMs are carefully aligned in cat's eye double-pass configuration to ensure that the laser beam direction does not change when the rf is tuned.

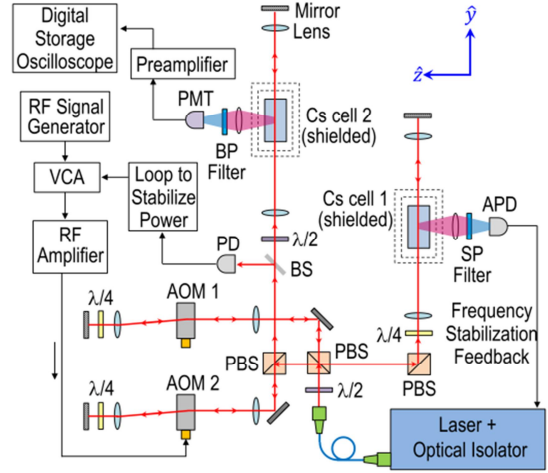

**FIG. 1S. Schematic diagram of the experimental setup.** AOM: acousto-optic modulator, PMT: photomultiplier tube, PD: photodiode, APD: avalanche PD, BS: beam sampler, PBS: polarizing beam splitter, BP: band-pass, SP: short-pass.

A lens of focal length 20 cm focuses the beam to a $1/e^2$ radius $r = 63 \pm 3$ μm and Rayleigh range 16 mm. The fluorescence collection lens system has a diameter of 25 mm and is placed ~ 67 mm from the excitation region. The 10-nm band-pass (BP) filter has center wavelength 670 nm (694 nm) for the $7d_{3/2}$ ($7d_{5/2}$) experiments and cannot resolve between the different decay channels shown in Fig. 1(a) of the main article. The laser power incident on Cs cell 2 is stabilized using a feedback loop that controls the rf power to AOM2. The Cs cells are heated to ~135°C (~100°C) for the $7d_{3/2}$ ($7d_{5/2}$) state measurements and are placed in two-layers of mu-metal shielding to reduce the residual magnetic field to ~2 mG.

## B. Additional comments on the experimental setup

As mentioned above, the frequency of the laser frequency is stabilized by locking it to a hyperfine line of the Doppler-free two-photon spectrum obtained in Cs cell 1. This ensures that the laser frequency does not drift during the experimental runs. Typical drifts are significantly below 5 kHz during an experimental run that takes several hours. The linewidth of the laser is ~100 kHz (obtained from the manufacturer data sheet).

To obtain the Doppler-free two-photon spectra from Cs cell 2 (as shown in Figs. 2 and 3 of the main article), we tune the laser frequency using AOM2 in cat's eye double pass configuration. The rf is supplied using a rf signal generator (SRS SG386, 1 μHz resolution). The tuning of the rf frequency is done at 1 Hz rate using a ramp generated from a waveform generator which controls the rf signal generator. Therefore, the frequency scans are perfectly linear (resulting in deviations much below 5 kHz during the frequency scan). We carefully align the setup to ensure than the double-passed laser beam does not show any deflection as the rf is tuned. Additionally, we focus the laser beam inside the Cs cell 2 using a pair of lenses (each with focal length $f$ = 20 cm) placed $2f$ apart which ensures that the incoming and the retro-reflected beams overlap at the focal point even if there are small unanticipated deflections of the laser beam as the AOM2 frequency is tuned. We ensure that the laser power is kept constant (within 0.2%) while the rf is tuned by implementing an electronic feedback loop that controls the rf power supplied to AOM2. These result in well-calibrated linear frequency scans with constant laser power.

The incident light is linearly polarized with a polarizing beam splitter (PBS) of extinction ratio 1000:1. With another PBS we measured that the Cs cell does not rotate the polarization and does not introduce any ellipticity. Heating the cell did not change these observations.

The fluorescence due to the $7d_J \to 6p_J$ decay is collected using a lens system of diameter 25 mm placed ~67 mm from the region where the laser beams are focussed. We place an aperture of 7 mm diameter located ~36 mm from the excitation region and this defines the solid angle $\Omega_C$ of light collection (and used as an input to the simulations). However, it should be noted that the focussed beam has a Rayleigh range of ~16 mm and the fluorescence region, as viewed from the detector, looks like a line rather than a point. Thus it is impossible to define a single point from which the fluorescence emanates. This introduces some errors in the definition of the solid angle. Nevertheless, we find that the measured intensity of the hyperfine lines match well with the simulated spectra suggesting that errors due to the definition of solid angle are small. Similarly, the finite Rayleigh range also introduces small errors in our definition of $\theta$. We could have allowed the solid angle and $\theta$ to be a parameter in our simulations (instead of fixing them to the measured values) but we found it to be unnecessary given the good agreement between the experiments and the simulations without free parameters.

**C. Comments on systematic errors**

The sources of systematic errors are the ac Stark shift, collisional shift, Zeeman shift due to the residual magnetic field and the second order Doppler shift.

The Cs cells are enclosed in two layers of mu-metal shield which reduce the magnetic field to < 2 mG. The mu-metal shield has wires wrapped around it in a toroidal fashion and can be degaussed in situ by passing ac current through the wires. Given that the residual magnetic field is low and that the experiments are performed with linearly polarized light (which eliminates optical pumping effects), the systematic errors due to Zeeman shifts are well below 1 kHz.

The second order Doppler shift $[= (v^2/c^2)\nu_L]$ is estimated to be ~0.5 kHz by considering the most probable speed $v$ of the atoms in the Cs cell.

We measure the collisional shift by recording spectra at several different temperatures ($T$) of the Cs cell 2. We convert the temperature ($T$) to the Cs vapour pressure $p$ using the expression [1]: $\log_{10} p = 2.881 + 4.165 - 3830/T$ and fit the frequency shift vs. $p$ data to a linear function (Fig. 2S). We find that the individual hyperfine lines ($F'$) shift with pressure, while keeping the hyperfine splitting constant within ~ 5 kHz. This implies that the collisional shift is same for all the hyperfine lines of a particular $7d_J$ state. The systematic error due to collisional shift (~ 5 kHz) is estimated from the scatter in the HFS data taken at different temperatures. From the slope of the plots in Fig. 2S, we determine the mean collisional shift for the $7d_{3/2}$ ($7d_{5/2}$) to be $-31.6 \pm 1.8$ ($-16.8 \pm 2.6$) kHz/mTorr.

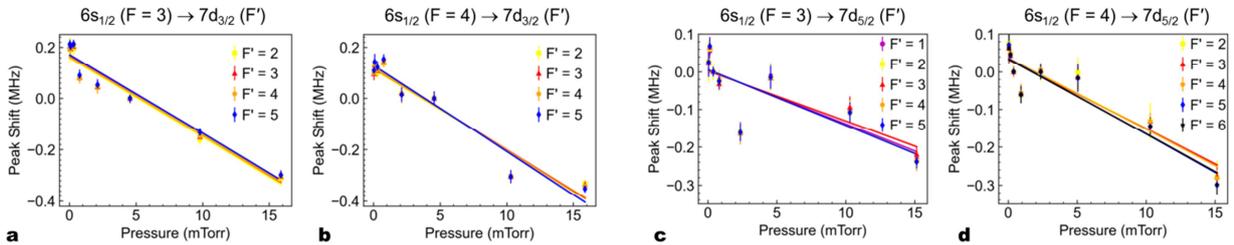

**Fig. 2S.** The shift in the peak position is plotted against the Cs vapour pressure. **(a, b)** $6s_{1/2} \to 7d_{3/2}$ transition. **(c, d)** $6s_{1/2} \to 7d_{5/2}$ transition. Shift in peak position of different hyperfine levels ($F'$) are measured by exciting atoms from the $6s_{1/2}$ ($F = 3$) and $6s_{1/2}$ ($F = 4$) levels. The shift is found to be independent of the state from which atoms are excited. It is observed that at each value of pressure, the data points for different $F'$ levels overlap, implying that all $F'$ levels shift by the same amount i.e. the HFS remains constant. All hyperfine levels of the $7d_{3/2}$ ($7d_{5/2}$) state have the same mean collisional shift $-31.6 \pm 1.8$ ($-16.8 \pm 2.6$) kHz/mTorr.

We measure the ac Stark shift by recording spectra at different powers ($P$) of the incident light. The frequency shift of a hyperfine line relative to data taken at the highest power is plotted against $P$ for the individual hyperfine lines ($F'$) in Fig. 3S. We find that the individual hyperfine lines ($F'$) shift with power while keeping the hyperfine splitting constant within ~ 3 kHz. This implies that the ac Stark shift is same for all the hyperfine lines of a particular $7d_J$ state. The systematic error due to ac Stark shift (~ 3 kHz) is estimated from the scatter in the HFS data taken at different powers. Considering the slope of the plots in Fig. 3S, the $1/e^2$ beam radius $r = 63 \pm 3$ μm and the 86% transmission through each end of the Cs vapour cell (resulting in bidirectional power of $1.5P$ at the center of the cell), we determine the mean ac Stark shift for the $7d_{3/2}$ ($7d_{5/2}$) to be $-53 \pm 5$ ($-55 \pm 9$) Hz/[W/cm$^2$]. These values agree well with those determined from theoretical calculations as discussed in section F.

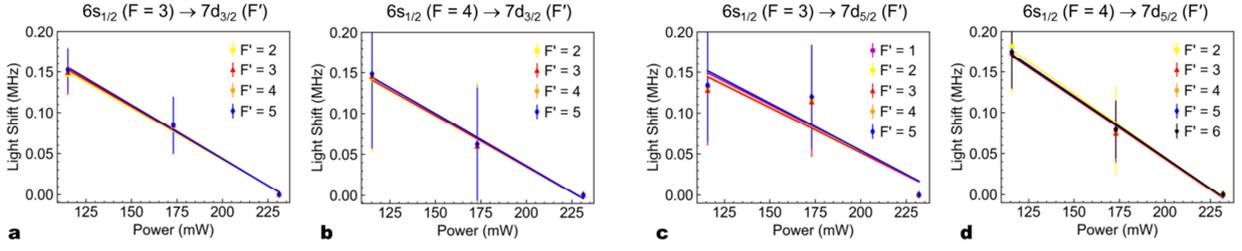

**Fig. 3S.** The shift in the peak position is plotted against the power incident on the Cs vapour cell. **(a, b)** $6s_{1/2} \to 7d_{3/2}$ transition. **(c, d)** $6s_{1/2} \to 7d_{5/2}$ transition. Shift in peak position of different hyperfine levels ($F'$) are measured by exciting atoms from the $6s_{1/2}$ ($F = 3$) and $6s_{1/2}$ ($F = 4$) levels. It is observed that at each value of power, the data points for different $F'$ levels overlap, implying that all $F'$ levels shift by the same amount i.e. the HFS remains constant. From the slope of the curves, we find that all hyperfine levels of the $7d_{3/2}$ ($7d_{5/2}$) state have the same light shift $-1.28 \pm 0.03$ ($-1.32 \pm 0.19$) MHz/W.

*Note on linewidths*: We determined the Lorentzian part of linewidths from the Voigt fits at different powers and different temperatures. The Lorentzian part of linewidths can be compared with theoretical values. We found that the measured Lorentzian linewidth does not change with power suggesting that ac Stark broadening is small. We found that the Lorentzian linewidth increases with pressure at ~ 95±7 (97±9) kHz/mTorr and has a zero-pressure intercept of ~ 2.1±0.1 (2.2±0.1) MHz which is slightly higher than the calculated natural linewidth ~1.73 (1.73) MHz of the $7d_{3/2}$ ($7d_{5/2}$) state [2]. The slightly higher value of the measured linewidth is attributed to the linewidth of the laser, transit time broadening and residual Doppler broadening arising from imperfect alignment.

### D. The measured hyperfine splitting

**Table 1S:** The hyperfine splitting (in MHz) measured in this work and a comparison with earlier experimental reports.

|  | Hyperfine Splitting | This work at θ = 53 ± 2° | Rahaman Ref. [3][a] | Rahaman Ref. [4][a] | Stalnaker Ref. [5][b] | Kumar Ref. [6] | Kortyna Ref. [7] | Lee Ref. [8] | Wang Ref. [9][c] |
|---|---|---|---|---|---|---|---|---|---|
| $7d_{3/2}$ | $\Delta_{54}$ | 36.7599(98) | 36.730(05) |  | 36.80(14) | 36.93(08) | 37.0(2) | 37.28(25) | 36.85(25) |
|  | $\Delta_{43}$ | 29.4272(52) | 29.429(13) |  | 29.60(08) | 29.59(08) | 29.2(2) | 30.21(35) | 29.70(25) |
|  | $\Delta_{32}$ | 22.0718(48) | 22.049(33) |  | 22.29(12) | 22.49(15) | 22.2(2) | 23.01(20) | 22.33(25) |
| $7d_{5/2}$ | $\Delta_{65}$ | 10.2817(28) |  | 10.225(09) | 10.39(28) |  |  | 10.39(35) | 10.20(25) |
|  | $\Delta_{54}$ | 8.5562(60) |  | 8.525(17) | 8.59(08) |  |  | 9.04(35) | 9.01(25) |
|  | $\Delta_{43}$ | 6.849(10) |  | 6.842(08) | 6.83(13) |  |  | 7.46(25) | 7.40(25) |
|  | $\Delta_{32}$ | 5.123(29) |  | 5.143(09) | 5.09(17) |  |  | 5.90(25) | 5.82(25) |
|  | $\Delta_{21}$ | 3.4013(96) |  | 3.320(07) | 3.38(15) |  |  |  |  |

[a] Average of data recorded for excitation from $F = 3$ and $F = 4$. [b] These values are obtained from the reported values of $A$ and $B$.

### E. Determination of hyperfine coupling constants (HCC) from the hyperfine splittings (HFS)

The expression relating the HFS to the HCCs depend on whether the calculations are performed to first-order or second-order in perturbation theory.

In first-order perturbation theory, the hyperfine interaction energy of a hyperfine level $F$ is given by [1,10,11]:

$$E_F = A\frac{K}{2} + B\frac{3K(K+1) - 4I(I+1)J(J+1)}{8I(2I-1)J(2J-1)} + C\frac{5K^2(K/4+1) + K[I(I+1) + J(J+1) + 3 - 3I(I+1)J(J+1)] - 5I(I+1)J(J+1)}{I(I-1)(2I-1)J(J-1)(2J-1)}$$

…(eq. 1)

where $A$, $B$ and $C$ are the magnetic dipole, the electric quadrupole and the magnetic octupole hyperfine coupling constants, respectively; $K = F(F+1) - I(I+1) - J(J+1)$ and $J$, $I$, $F$ are the electronic, nuclear and total angular momentum, respectively.

**$7d_{3/2}$ state:**

In first-order perturbation theory, the hyperfine splitting $\Delta_{ab}$ between any two hyperfine levels $F'_a$ and $F'_b$ is:

$\Delta_{54} = 5A + \frac{5}{7}B + \frac{40}{7}C$ …(eq. 2)

$\Delta_{53} = 9A + \frac{3}{7}B - \frac{48}{7}C$ …(eq.3)

$\Delta_{52} = 12A - \frac{2}{7}B + \frac{40}{7}C$ …(eq. 4)

$\Delta_{43} = 4A - \frac{2}{7}B - \frac{88}{7}C$ …(eq. 5)

$\Delta_{42} = 7A - B$ …(eq. 6)

$\Delta_{32} = 3A - \frac{5}{7}B + \frac{88}{7}C$ …(eq. 7)

**$7d_{5/2}$ state:**

In first-order perturbation theory, the hyperfine splitting $\Delta_{ab}$ between any two hyperfine levels $F'_a$ and $F'_b$ is:

$\Delta_{65} = 6A + \frac{18}{35}B + \frac{144}{35}C$ …(eq. 8)

$\Delta_{64} = 11A + \frac{11}{20}B - \frac{11}{35}C$ …( eq. 9)

$\Delta_{63} = 15A + \frac{9}{28}B - \frac{15}{7}C$ …( eq. 10)

$\Delta_{62} = 18A + \frac{24}{35}C$ …( eq. 11)

$\Delta_{54} = 5A + \frac{1}{28}B - \frac{31}{7}C$ …(eq. 12)

$\Delta_{53} = 9A - \frac{27}{140}B - \frac{219}{35}C$ …(eq. 13)

$\Delta_{52} = 12A - \frac{18}{35}B - \frac{24}{7}C$ …(eq. 14)

$\Delta_{43} = 4A - \frac{8}{35}B - \frac{64}{35}C$ …(eq. 15)

$\Delta_{42} = 7A - \frac{11}{20}B + C$ …(eq. 16)

$\Delta_{32} = 3A - \frac{9}{28}B + \frac{99}{35}C$ …(eq. 17)

$\Delta_{51} = 14A - \frac{4}{5}B + \frac{8}{5}C$ …(eq. 18)

$\Delta_{41} = 9A - \frac{117}{140}B + \frac{211}{35}C$ …(eq. 19)

$\Delta_{31} = 5A - \frac{17}{28}B + \frac{55}{7}C$ …(eq. 20)

$\Delta_{21} = 2A - \frac{2}{7}B + \frac{176}{35}C$ …(eq. 21)

We obtain the values of *A, B* and *C* of the $7d_{3/2}$ ($7d_{5/2}$) state by using the measured values of HFS reported in Table 1S as inputs to the left-hand-side of the equations 2-7 (8-21) and performing a global fitting to the equations. The obtained values of *A, B* and *C* are similar, i.e. within error bars, to the values reported in Table I of the main article which are obtained by considering the second order correction as described below.

To obtain more precise values of *A, B* and *C*, we consider the second-order correction reported in Ref. [12]. The corrections are smaller than the statistical errors in our experiment and therefore they do not change the results significantly. Nevertheless, we determine the values of *A, B* and *C* including the corrections and report them in Table I of the main article. The expressions for *A, B* and *C* including second-order correction are taken from Ref. [12] and reproduced below. The $\eta$ term gives the second-order correction due to the M1-M1 hyperfine interaction while the $\zeta$ term gives the second-order correction due to the M1-E2 hyperfine interaction.

**7$d_{3/2}$ state:**

$$A = \frac{11}{120}\Delta_{54} + \frac{2}{21}\Delta_{43} + \frac{3}{56}\Delta_{32} - \frac{1}{2520}\eta - \frac{1}{60}\zeta \qquad \text{where } -\frac{1}{2520}\eta = 7.38\times 10^{-2}\text{ kHz}, -\frac{1}{60}\zeta = -1.23\times 10^{-4}\text{ kHz}$$

$$B = \frac{77}{120}\Delta_{54} - \frac{1}{3}\Delta_{43} - \frac{5}{8}\Delta_{32} + \frac{1}{180}\eta - \frac{7}{120}\sqrt{\frac{1}{105}}\zeta \qquad \text{where } \mathbf{\frac{1}{180}\eta = -1.03\text{ kHz}}, -\frac{7}{120}\sqrt{\frac{1}{105}}\zeta = -4.30\times 10^{-4}\text{ kHz}$$

$$C = \frac{7}{480}\Delta_{54} - \frac{1}{24}\Delta_{43} + \frac{1}{32}\Delta_{32} + \frac{1}{480}\sqrt{\frac{1}{105}}\zeta \qquad \text{where } \frac{1}{480}\sqrt{\frac{1}{105}}\zeta = 1.54\times 10^{-5}\text{ kHz}$$

**7$d_{5/2}$ state:**

$$A = \frac{61}{1320}\Delta_{65} + \frac{267}{3080}\Delta_{54} + \frac{9}{280}\Delta_{43} + \frac{9}{280}\Delta_{32} + \frac{9}{280}\Delta_{21} + \frac{1}{3780}\eta - \frac{1}{210}\sqrt{\frac{1}{105}}\zeta$$

$$\text{where } \frac{1}{3780}\eta = 4.92\times 10^{-2}\text{ kHz}, -\frac{1}{210}\sqrt{\frac{1}{105}}\zeta = 3.51\times 10^{-5}\text{ kHz}$$

$$B = \frac{245}{264}\Delta_{65} + \frac{1}{88}\Delta_{54} - \frac{5}{8}\Delta_{43} - \frac{5}{8}\Delta_{32} - \frac{5}{8}\Delta_{21} + \frac{1}{54}\eta + \frac{1}{12}\sqrt{\frac{1}{105}}\zeta \quad \text{where } \mathbf{\frac{1}{54}\eta = 3.44\text{ kHz}}, \frac{1}{12}\sqrt{\frac{1}{105}}\zeta = -6.15\times 10^{-4}\text{ kHz}$$

$$C = -\frac{21}{352}\Delta_{65} + \frac{45}{352}\Delta_{54} - \frac{11}{352}\Delta_{43} - \frac{11}{352}\Delta_{32} - \frac{11}{352}\Delta_{21} + \frac{1}{48}\sqrt{\frac{1}{105}}\zeta \qquad \text{where } \frac{1}{48}\sqrt{\frac{1}{105}}\zeta = -1.54\times 10^{-4}\text{ kHz}$$

**F. Calculation of polarizability and ac Stark shift:**

The dynamic polarizability $\alpha(\omega,J)$ of an atom in a state with angular momentum $J$ interacting with a laser field of angular frequency $\omega$ is given by [13,14]:

$$\alpha(\omega,J) = \frac{2}{3(2J+1)} \sum_{J'} \frac{\omega_{J',J} |\langle J | d | J' \rangle|^2}{\omega_{J',J}^2 - \omega^2}$$

where $\langle J | d | J' \rangle$ is the dipole matrix element for $J \rightarrow J'$ transition and $\omega_{J',J}$ ($= \omega_{J'} - \omega_J$) is the resonant frequency for the transition. In the polarizability of the $6s_{1/2}$ state, all $6s_{1/2} \rightarrow np_{1/2}$ and $6s_{1/2} \rightarrow np_{3/2}$ transitions contribute to the sum.

**7$d_{3/2}$ state:** The relevant matrix elements and wavelengths are taken from Ref. [2] and listed in Table 2S. In the polarizability of the $7d_{3/2}$ state, all $7d_{3/2} \rightarrow np_{1/2}$, $7d_{3/2} \rightarrow np_{3/2}$ and $7d_{3/2} \rightarrow nf_{5/2}$ transitions contribute to the sum and the relevant transition energies and matrix elements are listed in Table 2S. The differential polarizability $\Delta\alpha$ [$= \alpha(7d_{3/2}) - \alpha(6s_{1/2})$] in atomic units is calculated and plotted in Fig. 4S(a) as a function of laser wavelength.

**7$d_{5/2}$ state:** The relevant matrix elements and wavelengths are taken from Ref. [2] and listed in Table 3S. In the polarizability of the $7d_{5/2}$ state, all $7d_{5/2} \rightarrow np_{3/2}$, $7d_{5/2} \rightarrow nf_{5/2}$ and $7d_{5/2} \rightarrow nf_{7/2}$ transitions contribute to the sum and the relevant transition energies and matrix elements are listed in Table 3S. The differential polarizability $\Delta\alpha$ [$= \alpha(7d_{5/2}) - \alpha(6s_{1/2})$] in atomic units is calculated and plotted in Fig. 4S(b) as a function of laser wavelength.

The ac Stark shift per unit laser intensity ($\delta f/I$) is related to $\Delta\alpha$ through the expression:

$$\delta f/I = -\Delta\alpha/2c\epsilon_0 h$$

where $h$ is Planck's constant, $c$ is the speed of light and $\epsilon_0$ is the permittivity of free space.

The calculated values of $\delta f/I$ are $-54.0$ Hz/(W/cm$^2$) and $-52.4$ Hz/(W/cm$^2$) for the $7d_{3/2}$ and $7d_{5/2}$ states, respectively.

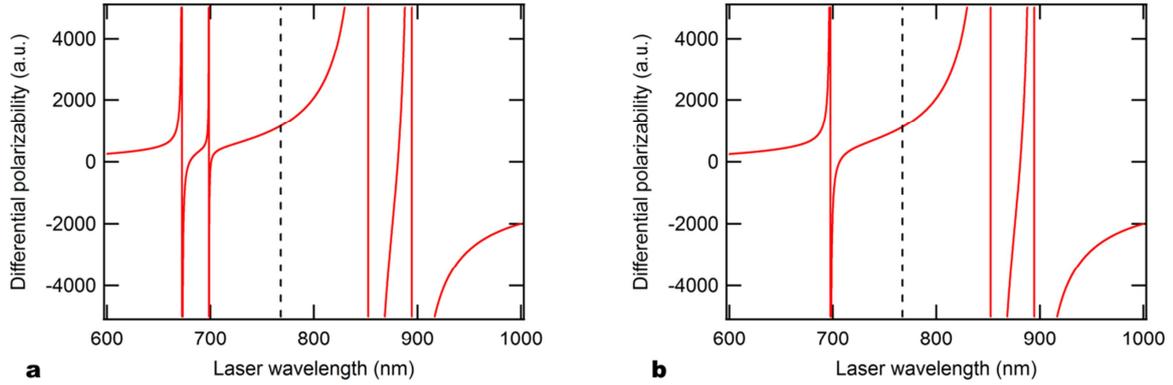

**Fig. 4S.** The calculated differential polarizability plotted as a function of the laser wavelength. The dashed line indicates the wavelength at which the ac Stark shift is measured in the experiment. **(a)** $\alpha(7d_{3/2}) - \alpha(6s_{1/2})$. **(b)** $\alpha(7d_{5/2}) - \alpha(6s_{1/2})$.

**Table 2S.** Values of transition energies and matrix elements [2] relevant to calculations for the $7d_{3/2}$ state. Note that the polarizability $\alpha(6s_{1/2})$ of the $6s_{1/2}$ state at 767.8 nm is determined almost entirely by the contributions from the $6s_{1/2} \rightarrow 6p_{1/2}$ and the $6s_{1/2} \rightarrow 6p_{3/2}$ transitions. On the other hand, the polarizability $\alpha(7d_{3/2})$ of the $7d_{3/2}$ state at 767.8 nm depends on contributions from several states.

| Transition | Energy (cm$^{-1}$) | $\langle J \mid d \mid J' \rangle$ (a.u.) | Contribution to $\alpha(6s_{1/2})$ at 767.8 nm (a.u.) | Transition | Energy (cm$^{-1}$) | $\langle J \mid d \mid J' \rangle$ in a.u. | Contribution to $\alpha(7d_{3/2})$ at 767.8 nm (a.u.) |
|---|---|---|---|---|---|---|---|
| $6s_{1/2} \rightarrow 6p_{1/2}$ | 11178.268 | 4.4978 | -370.3683 | $7d_{3/2} \rightarrow 6p_{1/2}$ | -14869.566 | 2.054 | -44.57 |
| $6s_{1/2} \rightarrow 6p_{3/2}$ | 11732.307 | 6.3349 | -1077.2327 | $7d_{3/2} \rightarrow 6p_{3/2}$ | -14315.527 | 0.9758 | -14.12 |
| $6s_{1/2} \rightarrow 7p_{1/2}$ | 21765.348 | 0.2781 | 0.4050 | $7d_{3/2} \rightarrow 7p_{1/2}$ | -4282.490 | 6.55 | 44.42 |
| $6s_{1/2} \rightarrow 7p_{3/2}$ | 21946.395 | 0.57417 | 1.6964 | $7d_{3/2} \rightarrow 7p_{3/2}$ | -4101.437 | 3.31 | 10.76 |
| $6s_{1/2} \rightarrow 8p_{1/2}$ | 25708.838 | 0.092 | 0.0324 | $7d_{3/2} \rightarrow 8p_{1/2}$ | -338.999 | 31.97 | 74.77 |
| $6s_{1/2} \rightarrow 8p_{3/2}$ | 25791.509 | 0.232 | 0.2049 | $7d_{3/2} \rightarrow 8p_{3/2}$ | -256.325 | 14.352 | 11.39 |
| $6s_{1/2} \rightarrow 9p_{1/2}$ | 27636.998 | 0.0429 | 0.0063 | $7d_{3/2} \rightarrow 9p_{1/2}$ | 1589.162 | 9.02 | -28.30 |
| $6s_{1/2} \rightarrow 9p_{3/2}$ | 27681.676 | 0.13 | 0.0574 | $7d_{3/2} \rightarrow 9p_{3/2}$ | 1633.844 | 3.564 | -4.55 |
| $6s_{1/2} \rightarrow 10p_{1/2}$ | 28726.816 | 0.0248 | 0.0020 | $7d_{3/2} \rightarrow 10p_{1/2}$ | 2678.978 | 2.86 | -4.93 |
| $6s_{1/2} \rightarrow 10p_{3/2}$ | 28753.678 | 0.086 | 0.0237 | $7d_{3/2} \rightarrow 10p_{3/2}$ | 2705.843 | 1.165 | -0.83 |
| $6s_{1/2} \rightarrow 11p_{1/2}$ | 29403.419 | 0.0162 | 0.0008 | $7d_{3/2} \rightarrow 11p_{1/2}$ | 3355.589 | 1.554 | -1.87 |
| $6s_{1/2} \rightarrow 11p_{3/2}$ | 29420.825 | 0.0627 | 0.0122 | $7d_{3/2} \rightarrow 11p_{3/2}$ | 3372.990 | 0.637 | -0.32 |
| $6s_{1/2} \rightarrow 12p_{1/2}$ | 29852.43 | 0.0115 | 0.0004 | $7d_{3/2} \rightarrow 12p_{1/2}$ | 3804.597 | 1.029 | -0.95 |
| $6s_{1/2} \rightarrow 12p_{3/2}$ | 29864.341 | 0.0486 | 0.0071 | $7d_{3/2} \rightarrow 12p_{3/2}$ | 3816.511 | 0.423 | -0.16 |
| $6s_{1/2} \rightarrow 13p_{1/2}$ | 30165.667 | 0.0087 | 0.0002 | $7d_{3/2} \rightarrow 13p_{1/2}$ | 4117.834 | 0.755 | -0.56 |
| $6s_{1/2} \rightarrow 13p_{3/2}$ | 30174.177 | 0.0392 | 0.0046 | $7d_{3/2} \rightarrow 13p_{3/2}$ | 4126.344 | 0.311 | -0.10 |
| $6s_{1/2} \rightarrow 14p_{1/2}$ | 30392.873 | 0.0069 | 0.0001 | $7d_{3/2} \rightarrow 14p_{1/2}$ | 4345.038 | 0.59 | -0.37 |
| $6s_{1/2} \rightarrow 14p_{3/2}$ | 30399.165 | 0.0326 | 0.0031 | $7d_{3/2} \rightarrow 14p_{3/2}$ | 4351.328 | 0.2429 | -0.06 |
| | | | | $7d_{3/2} \rightarrow 4f_{5/2}$ | -1575.607 | 13.03 | 58.55 |
| | | | | $7d_{3/2} \rightarrow 5f_{5/2}$ | 923.468 | 43.41 | -377.17 |
| | | | | $7d_{3/2} \rightarrow 6f_{5/2}$ | 2281.679 | 1.82 | -1.68 |
| | | | | $7d_{3/2} \rightarrow 7f_{5/2}$ | 3100.148 | 2.21 | -3.46 |
| | | | | $7d_{3/2} \rightarrow 8f_{5/2}$ | 3630.909 | 1.82 | -2.81 |
| | | | | $7d_{3/2} \rightarrow 9f_{5/2}$ | 3994.480 | 1.482 | -2.09 |
| | | | | $7d_{3/2} \rightarrow 10f_{5/2}$ | 4254.331 | 1.231 | -1.56 |
| | | | | $7d_{3/2} \rightarrow 11f_{5/2}$ | 4446.454 | 1.043 | -1.18 |
| | | | | $7d_{3/2} \rightarrow 12f_{5/2}$ | 4592.486 | 0.899 | -0.91 |

**Table 3S.** Values of transition energies and matrix elements [2] relevant to calculations for the $7d_{5/2}$ state. Note that the polarizability $\alpha(6s_{1/2})$ of the $6s_{1/2}$ state at 767.2 nm is determined almost entirely by the contributions from the $6s_{1/2} \rightarrow 6p_{1/2}$ and the $6s_{1/2} \rightarrow 6p_{3/2}$ transitions. On the other hand, the polarizability $\alpha(7d_{5/2})$ of the $7d_{5/2}$ state at 767.2 nm depends on contributions from several states.

| Transition | Energy (cm$^{-1}$) | $\langle J \| d \| J' \rangle$ (a.u.) | Contribution to $\alpha(6s_{1/2})$ at 767.2 nm (a.u.) | Transition | Energy (cm$^{-1}$) | $\langle J \| d \| J' \rangle$ in a.u. | Contribution to $\alpha(7d_{5/2})$ at 767.2 nm (a.u.) |
|---|---|---|---|---|---|---|---|
| $6s_{1/2} \rightarrow 6p_{1/2}$ | 11178.268 | 4.4978 | -368.1211 | $7d_{5/2} \rightarrow 6p_{3/2}$ | -14336.465 | 2.893 | -82.10 |
| $6s_{1/2} \rightarrow 6p_{3/2}$ | 11732.307 | 6.3349 | -1068.1240 | $7d_{5/2} \rightarrow 7p_{3/2}$ | -4122.368 | 9.640 | 61.10 |
| $6s_{1/2} \rightarrow 7p_{1/2}$ | 21765.348 | 0.2781 | 0.4053 | $7d_{5/2} \rightarrow 8p_{3/2}$ | -277.262 | 43.210 | 74.34 |
| $6s_{1/2} \rightarrow 7p_{3/2}$ | 21946.395 | 0.57417 | 1.6979 | $7d_{5/2} \rightarrow 9p_{3/2}$ | 1612.905 | 11.120 | -29.07 |
| $6s_{1/2} \rightarrow 8p_{1/2}$ | 25708.838 | 0.092 | 0.0324 | $7d_{5/2} \rightarrow 10p_{3/2}$ | 2684.904 | 3.610 | -5.24 |
| $6s_{1/2} \rightarrow 8p_{3/2}$ | 25791.509 | 0.232 | 0.2050 | $7d_{5/2} \rightarrow 11p_{3/2}$ | 3352.051 | 1.971 | -2.00 |
| $6s_{1/2} \rightarrow 9p_{1/2}$ | 27636.998 | 0.0429 | 0.0063 | $7d_{5/2} \rightarrow 12p_{3/2}$ | 3795.572 | 1.308 | -1.02 |
| $6s_{1/2} \rightarrow 9p_{3/2}$ | 27681.676 | 0.13 | 0.0574 | $7d_{5/2} \rightarrow 13p_{3/2}$ | 4105.405 | 0.961 | -0.60 |
| $6s_{1/2} \rightarrow 10p_{1/2}$ | 28726.816 | 0.0248 | 0.0020 | $7d_{5/2} \rightarrow 14p_{3/2}$ | 4330.390 | 0.750 | -0.39 |
| $6s_{1/2} \rightarrow 10p_{3/2}$ | 28753.678 | 0.086 | 0.0237 | $7d_{5/2} \rightarrow 4f_{5/2}$ | -1596.546 | 3.421 | 2.72 |
| $6s_{1/2} \rightarrow 11p_{1/2}$ | 29403.419 | 0.0162 | 0.0008 | $7d_{5/2} \rightarrow 4f_{7/2}$ | -1596.728 | 15.290 | 54.40 |
| $6s_{1/2} \rightarrow 11p_{3/2}$ | 29420.825 | 0.0627 | 0.0122 | $7d_{5/2} \rightarrow 5f_{5/2}$ | 902.530 | 11.661 | -17.70 |
| $6s_{1/2} \rightarrow 12p_{1/2}$ | 29852.43 | 0.0115 | 0.0004 | $7d_{5/2} \rightarrow 5f_{7/2}$ | 902.380 | 52.150 | -353.95 |
| $6s_{1/2} \rightarrow 12p_{3/2}$ | 29864.341 | 0.0486 | 0.0071 | $7d_{5/2} \rightarrow 6f_{5/2}$ | 2260.740 | 0.585 | -0.11 |
| $6s_{1/2} \rightarrow 13p_{1/2}$ | 30165.667 | 0.0087 | 0.0002 | $7d_{5/2} \rightarrow 6f_{7/2}$ | 2260.634 | 2.620 | -2.30 |
| $6s_{1/2} \rightarrow 13p_{3/2}$ | 30174.177 | 0.0392 | 0.0046 | $7d_{5/2} \rightarrow 7f_{5/2}$ | 3079.209 | 0.641 | -0.19 |
| $6s_{1/2} \rightarrow 14p_{1/2}$ | 30392.873 | 0.0069 | 0.0001 | $7d_{5/2} \rightarrow 8f_{7/2}$ | 3079.135 | 2.870 | -3.86 |
| $6s_{1/2} \rightarrow 14p_{3/2}$ | 30399.165 | 0.0326 | 0.0031 | $7d_{5/2} \rightarrow 9f_{5/2}$ | 3609.970 | 0.517 | -0.15 |
| | | | | $7d_{5/2} \rightarrow 9f_{7/2}$ | 3609.917 | 2.320 | -3.02 |
| | | | | $7d_{5/2} \rightarrow 10f_{5/2}$ | 3973.541 | 0.419 | -0.11 |
| | | | | $7d_{5/2} \rightarrow 10f_{7/2}$ | 3973.502 | 1.870 | -2.20 |
| | | | | $7d_{5/2} \rightarrow 11f_{5/2}$ | 4233.392 | 0.346 | -0.08 |
| | | | | $7d_{5/2} \rightarrow 11f_{7/2}$ | 4233.363 | 1.549 | -1.63 |
| | | | | $7d_{5/2} \rightarrow 12f_{5/2}$ | 4425.515 | 0.292 | -0.06 |
| | | | | $7d_{5/2} \rightarrow 12f_{7/2}$ | 4425.493 | 1.309 | -1.23 |

**G. Simulation of the Doppler-free two-photon spectra:**

The excitation and detection geometry relevant for the calculations is shown in Fig. 5S. The incoming laser propagating along $\hat{y}$ has linear polarization $\hat{\epsilon}_L$ (chosen along $\hat{z}$ without loss of generality), the detection direction is defined by the angle $\theta$ relative to $\hat{\epsilon}_L$, and the polarization $\hat{\epsilon}_s$ of the scattered light is decomposed into components along $\hat{\epsilon}_1$ and $\hat{\epsilon}_2$ which are orthogonal to each other.

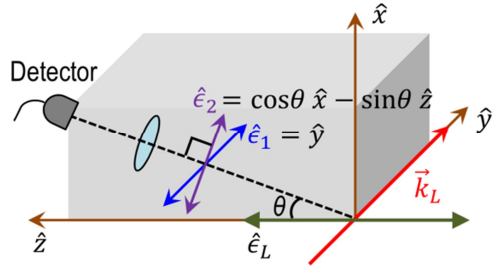

**Fig. 5S.** Excitation and detection geometry.

Following the formalism discussed in Refs. [15,16], and in absence of broadening mechanisms such as the transit time broadening, the scattering rate for a transition from state $|i\rangle$ to state $|f\rangle$ via all intermediate states $|e\rangle$, with emitted light of polarization $\hat{\epsilon}_s$ is given by:

$$R(i,f,\hat{\epsilon}_s,\hat{\epsilon}_L) \propto \int \left| \sum_e \frac{Q(\hat{\epsilon}_L,i,e)\langle e|\boldsymbol{d}.\hat{\epsilon}_s|f\rangle}{(\omega_{ef} - \omega_s - i\Gamma/2)} \right|^2 d\omega_s$$

Here, $\hat{\epsilon}_L$ is the polarization of the excitation laser and we choose this to be $(0,1,0)$ in the spherical basis $(e_-, e_0, e_+)$; $Q(\hat{\epsilon}_L,i,e)$ is the two-photon matrix element between $|i\rangle$ and $|e\rangle$; $\boldsymbol{d}$ is the dipole operator; $\omega_s$ is the frequency of the emitted photon and $\hat{\epsilon}_s$ is either $\hat{\epsilon}_2$ ($= \cos\theta\,\hat{x} - \sin\theta\,\hat{z}$ in the cartesian coordinate system which is $(\frac{\cos\theta}{\sqrt{2}}, -\sin\theta, \frac{-\cos\theta}{\sqrt{2}})$ in the spherical basis) or $\hat{\epsilon}_1$ ($= \hat{y}$ in the cartesian coordinate system which is $(\frac{i}{\sqrt{2}}, 0, \frac{i}{\sqrt{2}})$ in the spherical basis). In the spherical basis the dot product $\hat{\epsilon}_s.\boldsymbol{d} = \sum_q (-1)^q \epsilon_{-q} d_q$ and we have to compute the sum separately for the two possible scattered polarizations.

The vectors of matrix elements $\langle e|\boldsymbol{d}|f\rangle$ are given by:

$$\langle e|d_q|f\rangle = (-1)^{J(f)+L(e)+S+1}\sqrt{(2J(e)+1)(2J(f)+1)} \begin{Bmatrix} L(e) & J(e) & S \\ J(f) & L(f) & 1 \end{Bmatrix}$$

$$\times (-1)^{F(f)+J(e)+I+1}\sqrt{(2F(e)+1)(2F(f)+1)} \begin{Bmatrix} J(e) & F(e) & I \\ F(f) & J(f) & 1 \end{Bmatrix} (-1)^{(F(e)-M(e))} \begin{pmatrix} F(e) & 1 & F(f) \\ -M(e) & q & M(f) \end{pmatrix} \langle e||d||f\rangle$$

Here, the reduced matrix element $\langle e||d||f\rangle$ is independent of $J, F$ and is common to all terms for a particular transition.

The two-photon matrix elements $Q(\hat{\epsilon}_L,i,e)$ are given by:

$$Q(\hat{\epsilon}_L,i,e) = \sum_p \frac{\langle i|\boldsymbol{d}.\hat{\epsilon}_L|p\rangle\langle p|\boldsymbol{d}.\hat{\epsilon}_L|e\rangle}{\hbar(\omega_L - \omega_{pi})}$$

where the sum runs over all intermediate states $|p\rangle$. In practice, it is sufficient to consider only the intermediate $6p_{1/2}$ and $6p_{3/2}$ states since the detuning $\omega_L - \omega_{pi}$, which appears in the denominator, is much larger for other states and thus their contribution to the sum is small. For the $6s_{1/2} \to 7d_{5/2}$ two-photon transition, the only intermediate state that needs to be considered is the $6p_{3/2}$ state transition. For the $6s_{1/2} \to 7d_{3/2}$ two-photon transition, we must include both the $6p_{1/2}$ and $6p_{3/2}$ states as intermediate states. We ignore the hyperfine splitting when computing the denominator in the expression for $Q(\hat{\epsilon}_L,i,e)$, but the fine structure is significant and is included in the calculations. The single photon detuning of the laser from the intermediate $6p_{1/2}$ state $[\Delta_{1/2} = \omega_L - \omega(p_{1/2})]$ is approximately 55.3 THz, whereas that from the $6p_{3/2}$ state $[\Delta_{3/2} = \omega_L - \omega(p_{3/2})]$ is approximately 38.7 THz. Thus we use $\Delta_{3/2} = 0.7\,\Delta_{1/2}$ in the calculations.

The laser transit time broadening is included by considering that the atom passing through the laser beam experiences an intensity pulse $I_0 e^{-\gamma^2 t^2}$, the Fourier transform of which is defined following Ref. [15]:

$$\tilde{G}(\omega) = 2\int E_+(t)\,E_-(t)\,e^{2i\omega t}\,dt \propto 2\,(\sqrt{\pi}/\gamma)\,e^{-(\omega-\omega_L)^2/\gamma^2}$$

where $E_+(t)$ and $E_-(t)$ represent the fields travelling in opposite directions.

Substituting $\omega_s = 2\omega - \omega_{fi}$ into the expression for $R(i,f,\hat{\epsilon}_s,\hat{\epsilon}_L)$, we have:

$$R(i,f,\hat{\epsilon}_s,\hat{\epsilon}_L) \propto \int_{-\infty}^{\infty} \left| \sum_e \frac{Q(\hat{\epsilon}_L,i,e)\langle e|\boldsymbol{d}.\hat{\epsilon}_s|f\rangle}{(\omega_{ef} - 2\omega + \omega_{fi} - i\Gamma/2)} e^{-(\omega-\omega_L)^2/\gamma^2} \right|^2 d\omega$$

$$= \int_{-\infty}^{\infty} \left| \sum_e \frac{Q(\hat{\epsilon}_L,i,e)\langle e|\boldsymbol{d}.\hat{\epsilon}_s|f\rangle}{(\omega_{ei} - 2\omega - i\Gamma/2)} e^{-(\omega')^2/4\gamma^2} \right|^2 d\omega', \quad \text{where } \omega' = 2(\omega - \omega_L)$$

$$= \int_{-\infty}^{\infty} e^{-(\omega')^2/2\gamma^2} \left| \sum_e \frac{Q(\hat{\epsilon}_L,i,e)\langle e|\boldsymbol{d}.\hat{\epsilon}_s|f\rangle}{(\omega_{ei} - \omega' - 2\omega_L - i\Gamma/2)} \right|^2 d\omega'$$

$$= \int_{-\infty}^{\infty} \sum_{e,e'} e^{-(\omega')^2/2\gamma^2}\,\frac{Q(\hat{\epsilon}_L,i,e)\langle e|\boldsymbol{d}.\hat{\epsilon}_s|f\rangle}{(\Delta_e - \omega' - i\Gamma/2)}\,\frac{\{Q(\hat{\epsilon}_L,i,e')\langle e'|\boldsymbol{d}.\hat{\epsilon}_s|f\rangle\}^*}{(\Delta_{e'} - \omega' + i\Gamma/2)}\,d\omega', \text{where } \Delta_e = \omega_{ei} - 2\omega_L$$

$$= \sum_{e,e'} Q(\hat{\epsilon}_L, i, e) \langle e|\mathbf{d}.\hat{\epsilon}_s|f\rangle \{Q(\hat{\epsilon}_L, i, e') \langle e'|\mathbf{d}.\hat{\epsilon}_s|f\rangle\}^*$$

$$\times \frac{\pi \left( e^{\frac{(\Gamma - 2i\Delta_e)^2}{8\gamma^2}} \text{Erfc}\left[\frac{\Gamma - 2i\Delta_e}{2\sqrt{2}\gamma}\right] + e^{\frac{(\Gamma + 2i\Delta_{e'})^2}{8\gamma^2}} \text{Erfc}\left[\frac{\Gamma + 2i\Delta_{e'}}{2\sqrt{2}\gamma}\right] \right)}{\Gamma - i(\Delta_e - \Delta_{e'})}$$

$$= \sum_{e,e'} Q(\hat{\epsilon}_L, i, e) \langle e|\mathbf{d}.\hat{\epsilon}_s|f\rangle \{Q(\hat{\epsilon}_L, i, e') \langle e'|\mathbf{d}.\hat{\epsilon}_s|f\rangle\}^* \times \frac{\pi \left( f\left[\frac{-1}{i}\frac{\Gamma - 2i\Delta_e}{2\sqrt{2}\gamma}\right] + f\left[\frac{-1}{i}\frac{\Gamma + 2i\Delta_{e'}}{2\sqrt{2}\gamma}\right]\right)}{\Gamma - i(\Delta_e - \Delta_{e'})}$$

Here $f[z] = e^{-z^2}\text{Erfc}[-i\,z]$ is the Faddeeva W function, which can be conveniently evaluated in Wolfram Mathematica.

To simulate the spectrum, we calculate the sum assuming equal initial population in all the Zeeman sub-levels of the Cs hyperfine ground states $F = 3$ and $F = 4$. For the $7d_{5/2}$ state spectra, we consider the detected decay to the $6p_{3/2}$ state which comprises of the $F'' = 2,3,4,5$ hyperfine levels. For the $7d_{3/2}$ state spectra, we consider the detected decay to the $6p_{1/2}$ state which comprises of the $F'' = 3,4$ hyperfine levels. This gives an expression for the scattering detected in an infinitesimal solid angle around the detector direction $\theta$. We then need to correct for the solid angle $\Omega_c = 2\pi(1 - \cos\theta_c)$ of the circular collection lens which has half angle $\theta_c$. We express all the angular dependence in terms of the second Legendre polynomial, $P_2(\cos\theta)$, and then correct for the solid angle of the lens by making the replacement $P_2(\cos\theta) \to g(\theta_c)\,P_2(\cos\theta)$, where $g(\theta_c) = \cos\theta_c\,\cos^2(\theta_c/2)$ as derived in Ref. [16].